%% file: paper.tex
\documentclass[preprint]{aastex}

\usepackage{amsmath}
\usepackage{tikz}
\usetikzlibrary{shapes.geometric, arrows}
\usepackage[all]{hypcap}
\usepackage[utf8]{inputenc} 

\defcitealias{barnes_inference_2016}{Paper I}

\tikzstyle{box} = [rectangle, rounded corners, text centered, draw=black]
\tikzstyle{ghost} = [rectangle, rounded corners, text centered, draw=white]
\tikzstyle{arrow} = [thick, ->, >=stealth]

\begin{document}
	\title{Inference of Heating Properties from ``Hot'' Non-flaring Plasmas in Active Region Cores. II. Nanoflare Trains}
	\shorttitle{``Hot'' Non-flaring Plasmas II. Nanoflare Trains}
	\author{W. T. Barnes\altaffilmark{1}}
	\author{P. J. Cargill\altaffilmark{2,3}}
	\and
	\author{S. J. Bradshaw\altaffilmark{1}}
	\altaffiltext{1}{Department of Physics \& Astronomy, Rice University, Houston, TX 77251-1892; will.t.barnes@rice.edu, stephen.bradshaw@rice.edu}
	\altaffiltext{2}{Space and Atmospheric Physics, The Blackett Laboratory, Imperial College, London SW7 2BW; p.cargill@imperial.ac.uk}
	\altaffiltext{3}{School of Mathematics and Statistics, University of St Andrews, St Andrews, Scotland KY16 9SS}
	\begin{abstract}
	Despite its prediction over two decades ago, the detection of faint, high-temperature (``hot'') emission due to nanoflare heating in non-flaring active region cores has proved challenging. Using an efficient two-fluid hydrodynamic model, this paper investigates the properties of the emission expected from repeating nanoflares (a nanoflare train) of varying frequency as well as the separate heating of electrons and ions. If the emission measure distribution ($\mathrm{EM}(T)$) peaks at $T = T_m$, we find that $\mathrm{EM}(T_m)$ is independent of details of the nanoflare train, and $\mathrm{EM}(T)$ above and below $T_m$ reflects different aspects of the heating. Below $T_m$ the main influence is the relationship of the waiting time between successive nanoflares to the nanoflare energy. Above $T_m$ power-law nanoflare distributions lead to an extensive plasma population not present in a monoenergetic train. Furthermore, in some cases characteristic features are present in $\mathrm{EM}(T)$. Such details may be detectable given adequate spectral resolution and a good knowledge of the relevant atomic physics. In the absence of such resolution we propose some metrics that can be used to infer the presence of ``hot'' plasma.
	\end{abstract}
	\section{Introduction}
	\label{sec:intro}
	\par The concept of heating the solar corona by nanoflares, first proposed by \citet{parker_nanoflares_1988}, has been developed extensively over the past two decades \citep[e.g.][]{cargill_implications_1994,cargill_nanoflare_2004,klimchuk_solving_2006}. The term \textit{nanoflare} has now become synonomous with impulsive heating in the energy range $10^{24}-10^{27}$ erg, with no specific assumption regarding the underlying physical mechanism (for example, small-scale magnetic reconnection or hydromagnetic wave dissipation). In active region (AR) cores such as those we discuss in this paper, one strategy for constraining potential heating models is the analysis of the emission measure distribution as a function of temperature, $\mathrm{EM}(T)=\int n^2\mathrm{d}h$. \citet{cargill_implications_1994,cargill_nanoflare_2004} predicted that the $\mathrm{EM}(T)$ resulting from nanoflare heating should be wide, with a maximum value at $T=T_m\sim10^{6.5}$ K and have a faint, high-temperature (8-10 MK) component. Below $T_m$, there is a scaling $\mathrm{EM}(T)\sim T^a$ over a temperature range $10^6\lesssim T\lesssim T_m$, a result first discussed by \citet{jordan_structure_1975}. Observations from instruments onboard the \textit{Solar Dynamics Observatory} \citep[\textit{SDO},][]{pesnell_solar_2012} and \textit{Hinode} spacecraft \citep{kosugi_hinode_2007} have shown that $2\lesssim a\lesssim5$, with $T_m\approx10^{6.5-6.6}$ \citep{tripathi_emission_2011,warren_constraints_2011,warren_systematic_2012,winebarger_using_2011,schmelz_cold_2012,del_zanna_evolution_2015}.
	\par The emission component above $T_m$ has been the subject of less study, but is likely to be important as the so-called ``smoking gun'' of nanoflare heating since its properties may bear a close link to the actual heating. While many workers \citep{reale_evidence_2009,schmelz_hinode_2009,miceli_x-ray_2012,testa_hinode/eis_2012,del_zanna_elemental_2014,petralia_thermal_2014,schmelz_hot_2015} have claimed evidence of this hot, faint component of the emission measure, poor spectral resolution \citep{testa_temperature_2011,winebarger_defining_2012} and non-equilibrium ionization \citep{bradshaw_explosive_2006,reale_nonequilibrium_2008} may make a positive detection of nanoflare heating difficult. However, \citet{brosius_pervasive_2014} used observations from the \textit{EUNIS-13} sounding rocket to identify relatively faint emission from Fe XIX in a non-flaring active region (AR), suggesting temperatures of $\sim8.9$ MK.
	\par A scaling has been claimed for hot emission with $T>T_m$ such that $\mathrm{EM}\propto T^{-b}$, with $b>0$. This fit is usually done in the range $T_m\lesssim T\lesssim10^{7.2}$. However, measured values of these ``hotward" slopes are poorly constrained due to both the low magnitude of emission and the lack of available spectroscopic data in this temperature range \citep{winebarger_defining_2012}. \citet{warren_systematic_2012} find $7\lesssim b\lesssim10$ with uncertainties of $\pm2.5-3$, for 15 AR cores though \citet{del_zanna_elemental_2014}, using observations from the \textit{Solar Maximum Mission}, claim larger values for $b$. It must be noted though that reconstructing $\mathrm{EM}(T)$ from spectroscropic and narrow-band observations is non-trivial, with different inversion methods often giving significantly different results \citep{landi_monte_2012,guennou_can_2013}.
	\par An important parameter for any proposed coronal heating mechanism is the frequency of energy release along a single magnetic strand, where the observed loop comprises many such strands. Nanoflare heating can be classified as being either a \textit{high-} or \textit{low-frequency} (HF or LF, respectively). In the case of HF heating, $t_N$, the time between successive events, is such that $t_N\ll\tau_{cool}$, where $\tau_{cool}$ is a characteristic loop cooling time, and in the case of LF heating $t_N\gg\tau_{cool}$ \citep{mulu-moore_can_2011,warren_constraints_2011,bradshaw_diagnosing_2012,reep_diagnosing_2013,cargill_modelling_2015}. Steady heating is just HF heating in the limit $t_N\to0$. While a determination of $t_N$ is of great importance, its measurement is challenging. For example, while direct observations of possible reconnection-associated heating through short timescale changes in loop structure and emission is feasible, as demonsrated by the Hi-C rocket flight \citep{cirtain_energy_2013,cargill_solar_2013}, longer duration observations are required to constrain $t_N$. The previously-mentioned difficulties in reconstructing $\mathrm{EM}(T)$ must also be bourne in mind. Efforts to measure the heating frequency using narrow-band observations of intensity fluctuations in AR cores  have proved similarly difficult \citep{ugarte-urra_determining_2014}.
	\par The use of hydrodynamic loop models, combined with sophisticated forward modeling, is a useful method for assessing a wide variety of heating scenarios. Such models of nanoflare-heated loops have found emission measure slopes consistent with those derived from observations in the temperature range $T<T_m$. For example, while \citet{bradshaw_diagnosing_2012} found that the full range of $a$ could not be accounted for with low-frequency nanoflares, \citet{reep_diagnosing_2013} showed that using a tapered nanoflare train allowed for $0.9\lesssim a\lesssim4.5$. \citet{cargill_active_2014}, using a 0D loop model, investigated a large range of heating frequencies, $250\le t_N\le5000$ s, and found that only when $t_N$ was between a few hundred and 2000 seconds and proportional to the nanoflare energy could the full range of observed emission measure slopes be found.
	\par An analogous approach can be used to investigate the properties of the ``hot" coronal component expected from nanoflare heating, and is the subject of the present series of papers. In \citet{barnes_inference_2016} \citepalias[hereafter]{barnes_inference_2016}, we looked at the hot plasma properties due to a single isolated nanoflare. The effects of heating pulse duration, changes to conductive cooling due to heat flux limiting, differential heating of electrons and ions, and non-equilibrium ionization (NEI) were studied. It was shown that signatures of nanoflare heating are likely to be found in the temperature range $4\lesssim T\lesssim 10$ MK. The prospect of measurable signatures above $10$ MK was found to be diminished for short heating pulses (with duration $<100$ s), NEI, and differential heating of the ions rather than the electrons. It is important to stress for a single nanoflare that while the ``hot'' plasma is present, it cannot actually be detected.
	\par Single nanoflares are a good proxy for the LF heating scenario, but a study of nanoflare heating over a range of heating frequencies requires that we consider a ``train'' of nanoflares along a magnetic strand \citep{viall_patterns_2011,warren_constraints_2011,reep_diagnosing_2013,cargill_modelling_2015}. In this paper, we use an efficient two-fluid hydrodynamic model to explore the effect of a nanoflare train with varying $t_N$ on $\mathrm{EM}(T)$, in particular for $T>T_m$. Preferential species heating, NEI, power-law nanoflare distributions, and the effects of a variable $t_N$ between events are considered and an emission measure ratio metric, similar to that discussed in \citet{brosius_pervasive_2014}, is used to characterize the various results. \autoref{sec:methods} discusses the numerical model we have used to conduct this study and the parameter space we have investigated. \autoref{sec:results} shows the resulting emission measure distributions and diagnostics for the entire parameter space. Finally, \autoref{sec:conclusions} discusses how our results may be interpreted in the broader context of nanoflare heating and provides some concluding comments on our findings.
	\section{Methodology}
	\label{sec:methods}
	\subsection{Numerical Model}
	\label{subsec:numerics}
	\par Hydrodynamic models are excellent tools for computing field-aligned quantities in coronal loops. However, because of the small cell sizes needed to resolve the transition region and consequently small timesteps demanded by thermal conduction, the use of such models in large parameter space explorations is made impractical by long computational runtimes \citep{bradshaw_influence_2013}. We use the popular 0D enthalpy-based thermal evolution of loops (EBTEL) model \citep{klimchuk_highly_2008,cargill_enthalpy-based_2012,cargill_enthalpy-based_2012-1,cargill_modelling_2015} in order to efficiently simulate the evolution of a coronal loop over a large parameter space. This model, which has been successfully benchmarked against the 1D hydrodynamic HYDRAD code of \citet{bradshaw_influence_2013}, computes, with very low computational overhead, time-dependent, spatially-averaged loop quantities.
	\par In order to treat the evolution of the electron and ion populations separately, we use a modified version of the usual EBTEL equations. This amounts to computing spatial averages of the two-fluid hydrodynamic equations over both the transition region and corona\footnote{The two-fluid EBTEL  code is freely available and can be downloaded at: \url{https://github.com/rice-solar-physics/ebtelPlusPlus}.}. A full description and derivation of these equations can be found in Appendix B of \citetalias{barnes_inference_2016}. The relevant two-fluid pressure and density equations are,
	\begin{align}
		\frac{d}{dt}\bar{p}_e &= \frac{\gamma - 1}{L}\left\lbrack\psi_{TR} - (\mathcal{R}_{TR} + \mathcal{R}_C)\right\rbrack + k_B\bar{n}\nu_{ei}(\bar{T}_i - \bar{T}_e) + (\gamma - 1)\bar{Q}_e,\label{eq:ebtel2fl_pe}\\
		\frac{d}{dt}\bar{p}_i &= -\frac{\gamma - 1}{L}\psi_{TR} + k_B\bar{n}\nu_{ei}(\bar{T}_e - \bar{T}_i) + (\gamma - 1)\bar{Q}_i,\label{eq:ebtel2fl_pi}\\
		\frac{d}{dt}\bar{n} &= \frac{c_2(\gamma - 1)}{c_3\gamma Lk_B\bar{T}_e}\left(\psi_{TR} - F_{ce,0} - \mathcal{R}_{TR}\right),\label{eq:ebtel2fl_n}
	\end{align}
	where $c_2=\bar{T}_e/T_{e,a}\approx0.9$, $c_3=T_{e,0}/T_{e,a}\approx0.6$, $\nu_{ei}$ is the electron-ion binary Coulomb collision frequency, and $\psi_{TR}$ is a term included to maintain charge and current and neutrality. Additionally, $c_1=\mathcal{R}_{TR}/\mathcal{R}_C$ and its formulation is discussed in \citet{cargill_enthalpy-based_2012} with additional modifications detailed in Appendix A of \citetalias{barnes_inference_2016}. These equations are closed by the equations of state $p_e=k_BnT_e$ and $p_i=k_BnT_i$. In the cases where we treat the plasma as a single-fluid, we use the original EBTEL model as described in \citet{klimchuk_highly_2008,cargill_enthalpy-based_2012}.
	\par The loop is heated by a prescribed heating function, applied to either the electrons ($\bar{Q}_e$) or the ions ($\bar{Q}_i$). Both species cool through a combination of thermal conduction ($F_{ce,0},\,F_{ci,0}$) and an enthalpy flux to the lower atmosphere, with the electrons also undergoing radiative cooling ($\mathcal{R_C}$). In the case of conductive cooling, a flux limiter, $F_s=(1/2)fnk_BTV_e$, is imposed to mitigate runaway cooling in a low-density, high-temperature plasma. In all cases we use a saturation limit of $f=1$. See \citetalias{barnes_inference_2016} for a discussion of how $f$ is likely to affect the presence of hot emission in a nanoflare-heated plasma.
	\subsection{Energy Budget}
	\label{subsec:params}
	\begin{figure}
		\plotone{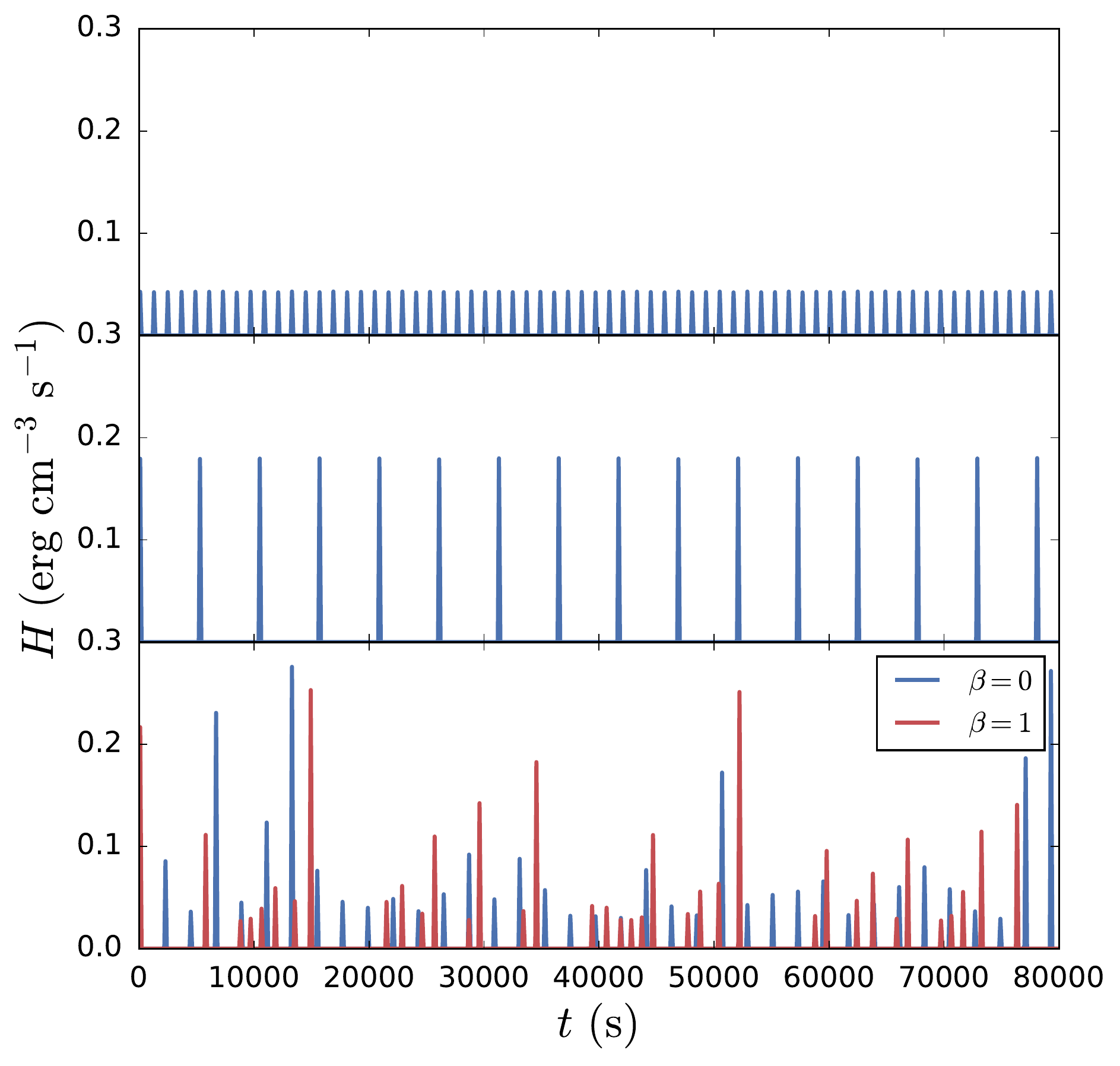}
		\caption{Examples of four different heating functions: uniform heating amplitudes for $t_N=1000$ s (top), uniform heating amplitudes for $t_N=5000$ s (middle), and heating amplitudes drawn from a power-law distribution with index $\alpha=-1.5$ (bottom). In the bottom panel, the events shown in red have waiting times that depend on the previous event energy ($\beta=1$) while the events shown in blue have uniform waiting times ($\beta=0$). The average waiting time in both cases is $t_N=2000$ s.}
		\label{fig:heating_funcs}
	\end{figure}
	\par We define our heating function in terms of a series of discrete heating events plus a static background heating to ensure that the loop does not drop to unphysically low temperatures and densities between events. For a triangular heating pulse of duration $\tau$ injected into a loop of half-length $L$ and cross-sectional area $A$, the total  event energy is $\varepsilon=LAH\tau/2$, where $H$ is the heating rate. Each model run will consist of $N$ heating events, each with peak amplitude $H_i$, and a steady background heating of $H_{bg}=3.5\times10^{-5}$ erg cm$^{-3}$ s$^{-1}$.
	\par Recent observations have suggested that loops in AR cores are maintained at an equilibrium temperature of $T_{m}\approx4$ MK \citep{warren_constraints_2011,warren_systematic_2012}. Using our modified two-fluid EBTEL model, we have estimated the time-averaged volumetric heating rate needed to keep a loop of half-length $L=40$ Mm at $\bar{T}\approx4$ MK as  $H_{eq}\sim3.6\times10^{-3}$ erg cm$^{-3}$ s$^{-1}$. In the single-fluid EBTEL model, this value is slightly lower because losses due to electron-ion collisions are ignored. Thus, to maintain an emission measure peaked about $T_{m}$, for triangular pulses, the individual event heating rates are constrained by,
	\begin{equation}
		\label{eq:heating_rate_constraint}
		H_{eq} = \frac{1}{t_{total}}\sum_{i=1}^N\int_{t_i}^{t_i+\tau}\mathrm{d}t~Q(t) = \frac{\tau}{2t_{total}}\sum_{i=1}^NH_i,
	\end{equation}
	where $t_{total}$ is the total simulation time. Note that if $H_i=H_0$ for all $i$, the heating rate for each event is $H_i=H_0=2t_{total}H_{eq}/N\tau$. Thus, for $L=40$ Mm and cross-sectional area $A=10^{14}$ cm$^2$, the average energy per event for a loop heated by $N=20$ nanoflares in $t_{total}=8\times10^4$ s is $\varepsilon=LAt_{total}H_{eq}/N\approx5.8\times10^{24}$ erg.
	\par We define the heating frequency in terms of the waiting time, $t_N$, between successive heating events. Following \citet{cargill_active_2014}, the range of waiting times is $250\le t_N\le5000$ s in increments of 250 s, for a total of 20 different possible heating frequencies. Additionally, $t_N$ can be written as $t_N=(t_{total}-N\tau)/N$, where we fix $t_{total}=8\times10^4$ s and $\tau=200$ s. Note that because $t_{total}$ and $\tau$ are fixed, as $t_N$ increases, $N$ decreases. Correspondingly, $\varepsilon_i=LA\tau H_i/2$, the energy injected per event, increases according to \autoref{eq:heating_rate_constraint} such that the total energy injected per run is constant.
	\par According to the nanoflare heating model of \citet{parker_nanoflares_1988}, turbulent loop footpoint motions twist and stress the field, leading to a buildup and subsequent release of energy. Following \citet{cargill_active_2014}, we let $\varepsilon_i\propto t_{N,i}^{\beta}$, where $\varepsilon_i,t_{N,i}$ are the total energy of event $i$ and waiting time following event $i$, respectively, and $\beta=1$ such that the event energy scales linearly with the waiting time. The reasoning for such an expression is as follows. Bursty, nanoflare heating is thought to arise from the stressing and subsequent relaxation of the coronal field. If a sufficient amount of energy is released into the loop, the field will need enough time to ``wind up'' again before the next event such that the subsequent waiting time is large. Conversely, if only a small amount of energy is released, the field will require a shorter re-winding time, resulting in a shorter interval between the subsequent events. Thus, this scaling provides a way to incorporate a more physically motivated heating function into a hydrodynamic model which cannot self-consistently determine the heat input based on the evolving magnetic field. \autoref{fig:heating_funcs} shows the various heating functions used for several example $t_N$ values. Note that when $\beta=1$, $t_N$ is the \textit{average} waiting time between events since the waiting time after a particular event $i$ is dependent on the energy of that event.
	\subsection{Heating Statistics}
	\label{subsec:heating_stats}
	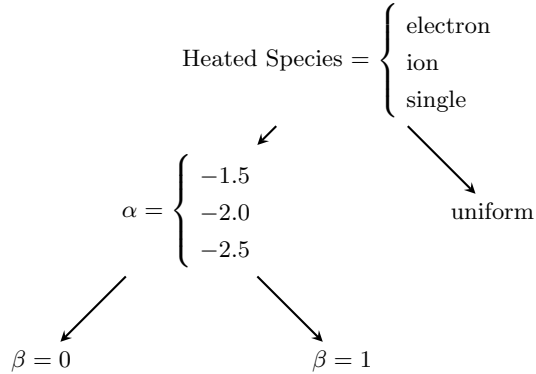
\begin{figure}
		\centering
		\input{parameter_space_graphic}
		\caption{Total Parameter space covered. ``single'' indicates a single-fluid model. $\alpha$ is the power-law index and $\beta$ indicates the scaling in the relationship $\varepsilon\propto T_N^{\beta}$, where $\beta=0$ corresponds to the case where $t_N$ and the event energy are independent (i.e. $t_N$ is uniform). Note that $(3~\alpha~\mathrm{values})\times(2~\beta~\mathrm{values})+\mathrm{uniform~heating}=$ 7 different types of heating functions per heated species.}
		\label{fig:parameter_space}
	\end{figure}
	\par We compute the peak heating rate per event in two different ways: 1) the heating rate is uniform such that $H_i=H_0$ for all $i$ and 2) $H_i$ is chosen from a power-law distribution with index $\alpha$ where $\alpha=-1.5,-2.0,$ or $-2.5$. For the second case, it should be noted that, when $t_N\approx5000$ s, $N\sim16$ events, meaning the events from a single run do not accurately represent the distribution of index $\alpha$. Thus, a sufficiently large number of runs, $N_{R}$, are computed for each $t_N$ to ensure that the total number of events is $N_{tot}=N\times N_{R}\sim10^4$ such that the distribution is well-represented. \autoref{fig:parameter_space} shows the parameter space we will explore. For each set of parameters and waiting time $t_N$, we compute the resulting emission measure distribution for $N$ events in a period $t_{total}$. This procedure is repeated $N_R$ times until $N\times N_R\sim10^4$ is satisfied. Thus, when $t_N=5000$ s and $N\sim16$, $N_R=625$, meaning the model is run 625 times with a waiting time of $t_N=5000$ s in order to properly fill out the event energy distribution.
	\subsection{Non-equilibrium Ionization}
	\label{subsec:nei}
	\par When considering the role of nanoflares in the production of hot plasma in AR cores, it is important to take non-equilibrium ionization (NEI) into account \citep{bradshaw_explosive_2006,reale_nonequilibrium_2008,barnes_inference_2016}. In a steady heating scenario, the ionization state is an adequate measure of the electron plasma temperature. Because the heating timescale is long (effectively infinite), the ionization state has plenty of time to come into equilibrium with the electron temperature.
	\par In a nanoflare train, when the heating frequency is high, the loop is not allowed to drain or cool sufficiently between events, meaning the ionization state is kept at or near equilibrium. However, as the heating frequency decreases, the loop is allowed to cool and drain more and more during the inter-event period. If the heating occurs on a short enough timescale, the ionization state will not be able to reach equilibrium with the electron plasma before the loop undergoes rapid cooling by thermal conduction. Furthermore, if the frequency is sufficiently low so as to allow the loop to drain during the inter-event period, the ionization equilibrium timescale will increase. Thus, in the context of intermediate- to low-frequency nanoflares, NEI should be considered.
	\par As in \citetalias{barnes_inference_2016}, we use the numerical code\footnote{This code has been made freely available by the author and can be downloaded at: \url{https://github.com/rice-solar-physics/IonPopSolver}.} outlined in \citet{bradshaw_numerical_2009} to asses the impact of NEI on our results. Given a temperature and density profile from EBTEL, we compute the non-equilibrium ionization states for Fe IX through XXVII and the corresponding effective electron temperature, $T_{eff}$, that would be inferred by assuming ionization equilibrium. Using $T_{eff}$, we are then able to compute a corresponding NEI emission measure distribution, $\mathrm{EM}(T_{eff})$.
	\section{Results}
	\label{sec:results}
	\par We now show the results of our nanoflare train simulations for each point in our multidimensional parameter space: species heated (single-fluid, electron or ion), power-law index ($\alpha$), waiting time ($t_N$), and waiting-time/event energy relationship ($\beta$). In each 0D hydrodynamic simulation, a loop of half-length $L=40$ Mm is heated by $N$ triangular events of duration $\tau=200$ s and peak heating rate $H_i$ for a duration of $t_{total}=8\times10^4$ s. The average interval between subsequent events is $t_N$ (in the uniform and $\beta=0$ cases, $t_{N,i}=t_N$ exactly for all $i$). We focus primarily on the emission measure distribution, $\mathrm{EM}(T)$, and observables typically calculated from $\mathrm{EM}(T)$. In all cases, the coronal emission measure is calculated according to the method outlined in section 3 of \citetalias{barnes_inference_2016}. The corresponding NEI results, $\mathrm{EM}(T_{eff})$, are calculated similarly, but using $T_{eff}$ (see \autoref{subsec:nei}) instead of $T$.
	\subsection{Emission Measure Distributions}
	\label{subsec:em_dist}
	\begin{figure*}[t]
		\plotone{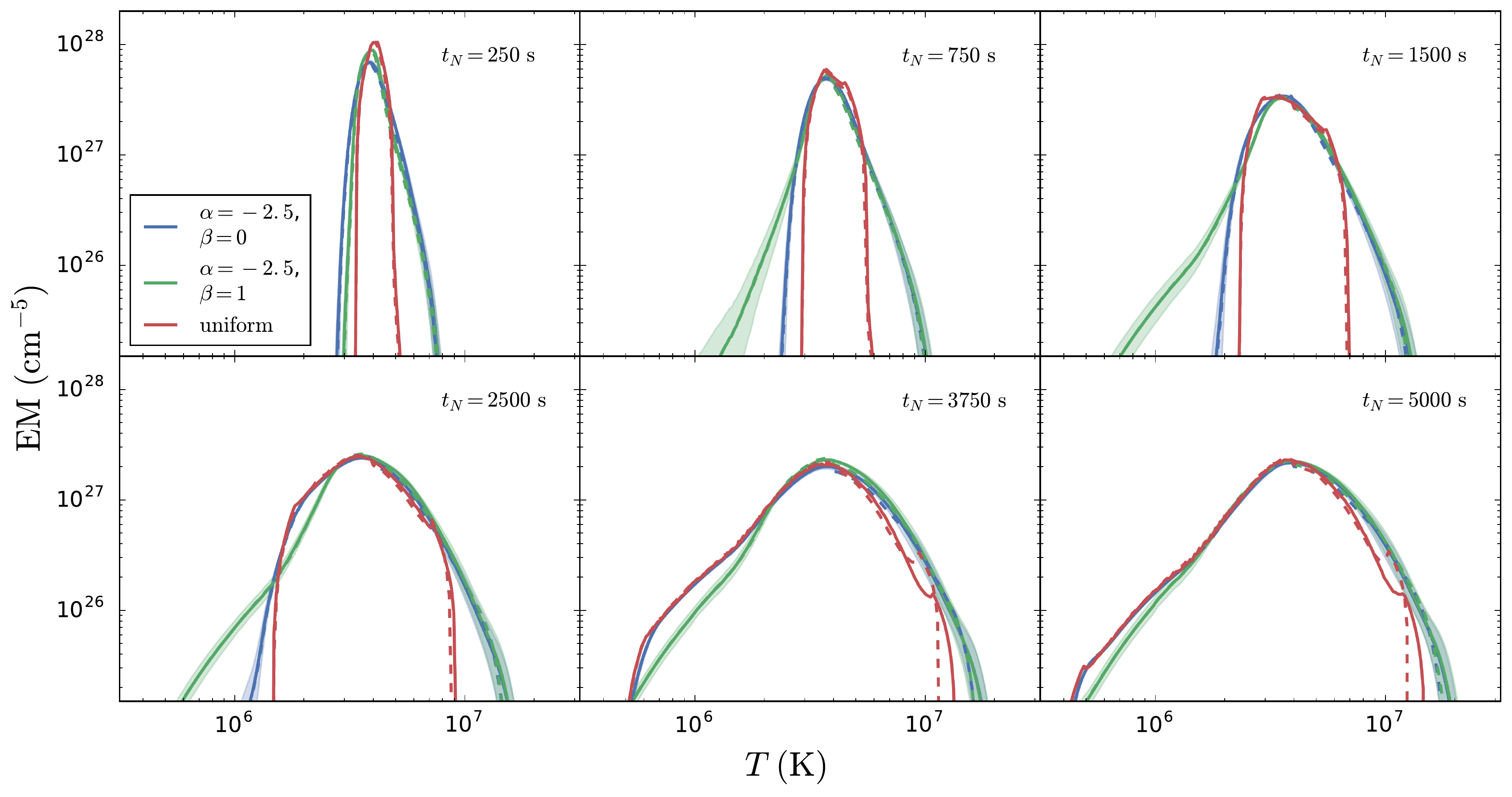}
		\caption{Emission measure distributions for waiting times $t_N=250,750,1500,2500,3750,5000$ s in the single-fluid case. The three types of heating functions shown are uniform heating rates (red), heating rates chosen from a power-law distribution of $\alpha=-2.5$ (blue), and heating rates chosen from a power-law distribution of $\alpha=-2.5$ where the waiting time after each event is proportional to the heating rate of the event (green). For the last case ($\beta=1$), $t_N$ is the average waiting time for all events. Note that in some panels, the blue $\beta=0$ curves may not be visible because they overlap heavily with the green $\beta=1$ curves. The solid lines in the two power law cases show the mean $\mathrm{EM}(T)$ over $N_R$ runs and the shading indicates one standard deviation from the mean. The dashed lines denote the corresponding $\mathrm{EM}(T_{eff})$ distribution. The standard deviation is not included in the NEI results.}
		\label{fig:single_em}
	\end{figure*}
	\begin{figure*}[t]
		\plotone{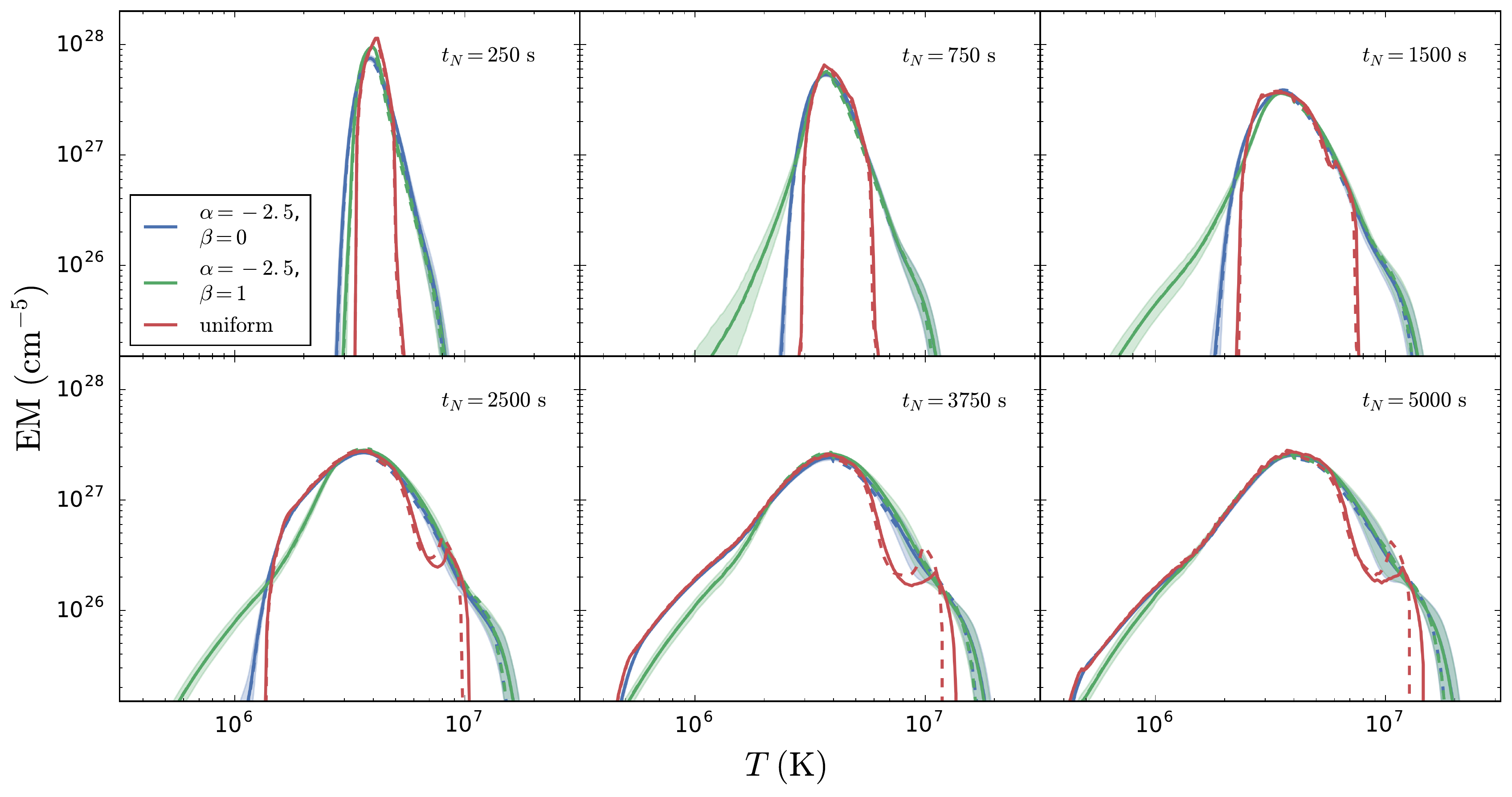}
		\caption{Same as \autoref{fig:single_em}, but for the case where only the electrons are heated.}
		\label{fig:el_em}
	\end{figure*}
	\begin{figure*}
		\plotone{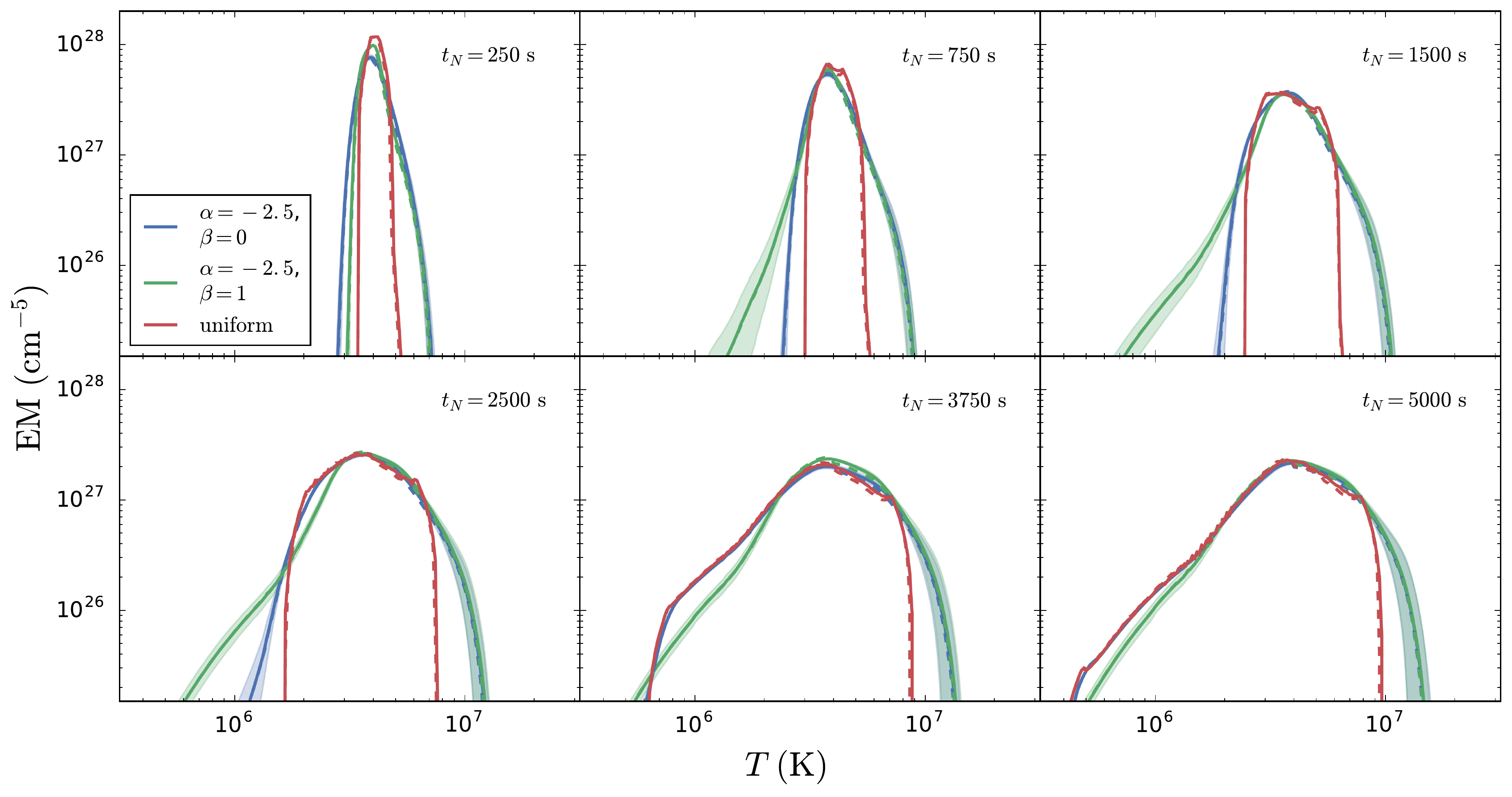}
		\caption{Same as \autoref{fig:single_em}, but for the case where only the ions are heated.}
		\label{fig:ion_em}
	\end{figure*}
	\par We compare $\mathrm{EM}(T)$ for three different types of heating functions, across a sample of six different heating frequencies. \autoref{fig:single_em}, \autoref{fig:el_em}, and \autoref{fig:ion_em} show the emission measure distributions for the single-fluid, electron heating, and ion heating cases respectively. Each panel corresponds to a different average waiting time ($t_N$) and includes three different types of heating functions: uniform heating events (red), events chosen from a single power-law distribution of index $\alpha=-2.5$ with constant waiting time ($\beta=0$ case, blue), and a waiting time that depends on the energy released in the preceding event ($\beta=1$ case, green). The dashed lines denote the corresponding NEI cases, $\mathrm{EM}(T_{eff})$. The cases shown in \citetalias{barnes_inference_2016} correspond approximately to the red curves in the lower right panels ($t_N=5000$ s) in each of the three figures since the loop is allowed to cool and drain completely before reheating and a single nanoflare energy is used.
	\par In \autoref{subsec:heating_stats}, we noted that for heating functions using a power-law energy distribution, for each $t_N$, we run the model $N_R$ times. Thus, for each point in our parameter space, we produce $N_R$ $\mathrm{EM}(T)$ curves. In order to present our results compactly, the solid lines in \autoref{fig:single_em}, \autoref{fig:el_em}, and \autoref{fig:ion_em} each show the mean $\mathrm{EM}(T)$ over all $N_R$ curves. The shading represents one standard deviation from the mean. In this way, we account for the variations that may occur because of a lack/excess of strong heating events due to limited sampling from the power-law distribution of possible heating rates.
	\par In all cases, $\mathrm{EM}(T)$ has some generic properties. Firstly, as expected from \citet{cargill_active_2014}, as $t_N$ increases, $\mathrm{EM}(T)$ widens, extending to both cooler ($<4$ MK) and hotter ($>4$ MK) temperatures. Secondly, for a prescribed value of $t_N$, the values of $\mathrm{EM}(T_m)$ and $T_m$ are approximately the same for all forms of heating. This can be attributed to the effective coupling between the species at $n(T_m)$ and suggests that $\mathrm{EM}(T)$ above and below $T_m$ can be considered separately, with each providing information about different aspects of the heating.
	\par For $T<T_m$, the extension of $\mathrm{EM}(T)$ toward cooler temperatures arises because as $t_N$ increases there is more time between successive heating events so that the loop cools to lower temperatures before being reheated. The dependence on $\alpha$ and $\beta$ is similar to that described in \citet{cargill_active_2014}. For both uniform heating (red) and a power-law index without waiting time (blue), $\mathrm{EM}(T)$ falls off more rapidly than when a waiting time is included (green) for intermediate frequencies. For example, for $t_N=1500$ s, both the uniform and $\beta=0$ cases show little to no emission below 2 MK while the $\beta=1$ case extends to temperatures well below 1 MK. Thus this part of $\mathrm{EM}(T)$ has information about the need for a waiting time, but not about the details of which species is heated.
	\par The behavior of $\mathrm{EM}(T)$ above $T_m$ is more complicated, but because $\mathrm{EM}(T_m)$ and $T_m$ are the same for all parameters, we can make a meaningful comparison between the different heating models. For the single-fluid model and short $t_N$, the emission measure distribution falls off sharply on the hot side for a uniform nanoflare train, but choosing heating events from a power law leads to a broader distribution. This just reflects the different initial temperatures generated with a power-law distribution since $T \simeq H^{2/7}$, where $H$ is the heating rate. As $t_N$ increases, the distribution for uniform heating gradually broadens as the initial temperature rises due to the lower density in which the heating occurs. A similar broadening occurs for power law heating distributions with the $\beta=0$ and $\beta=1$ results showing little difference. Note that the $\beta=0$ curve is barely visible as it overlaps almost completely with the $\beta=1$ curve. Especially interesting in this case are the results with NEI included. For a uniform nanoflare train, NEI plays no role up to $t_N$ = 2500 sec, but above that it restricts the temperatures that can be detected, as shown in \citetalias{barnes_inference_2016}. This hot emission is relocated to cooler temperatures, resulting in a small ``bump'' in the emission measure distribution near 10 MK. On the other hand, NEI plays almost no role in the power-law distributions, for either the $\beta=0$ or $\beta=1$ cases.
	\par For electron heating, the $\mathrm{EM}(T)$ curves for the different types of heating functions shown in \autoref{fig:el_em} evolve similarly to those shown in \autoref{fig:single_em}. For $t_N\le750$ s, the electron and single-fluid cases are quite similar at $T>T_m$. However, for $t_N\ge1500$ s $\mathrm{EM}(T)$ steepens just above 4 MK and then flattens out near 10 MK. This change in shape is most obvious in the uniform heating case where a distinct ``hot shoulder'' forms just above 10 MK. In the power law cases, this feature is less pronounced though $\mathrm{EM}(T)$ extends to slightly higher temperatures. Again for a power-law distribution NEI is not important while for uniform heating, the hot emission is again truncated and leads to a``bump'' in $\mathrm{EM}(T)$ near $10$ MK.
	\par When only the ions are heated (\autoref{fig:ion_em}), for intermediate to low heating frequencies (i.e. $t_N\ge1500$ s), $\mathrm{EM}(T)$ in the uniform heating case is truncated below 10 MK and in the power law cases extends to just above 10 MK for the longest waiting time ($t_N=5000$ s). This cutoff at lower temperatures is due to the fact that the electrons cannot ``see'' the ions until they have cooled well below their peak temperature. This is discussed in \citetalias{barnes_inference_2016} though in the single-pulse cases, the cutoff occured at lower temperatures. Additionally, in both the uniform and power law cases, the peak of $\mathrm{EM}(T)$ is wider for these low frequencies compared to those shown in the lower right panels of \autoref{fig:single_em} and \autoref{fig:el_em}. NEI now plays no role in any of the cases.
	\par From these results, we see that new information about the heating is potentially available above $T_m$, but unlike at lower temperatures (i.e. $T<T_m$), information about the role of a waiting time is lost. For high frequency nanoflares, there is no plasma above $10^{6.8}$ K in any of the heating scenarios. For intermediate heating frequencies, there is a significant enhancement in $\mathrm{EM}(T>T_m)$ for the power law cases relative to the uniform heating case in the single-fluid, electron heating, and ion heating cases. For low frequencies, this discrepancy is less pronounced, though the uniform single-fluid and electron heating cases show distinctive features in $\mathrm{EM}(T)$ near 10 MK.
	\subsection{Pre-nanoflare Density}
	\label{subsec:pre_nanoflare_density}
	\begin{figure*}
		\plotone{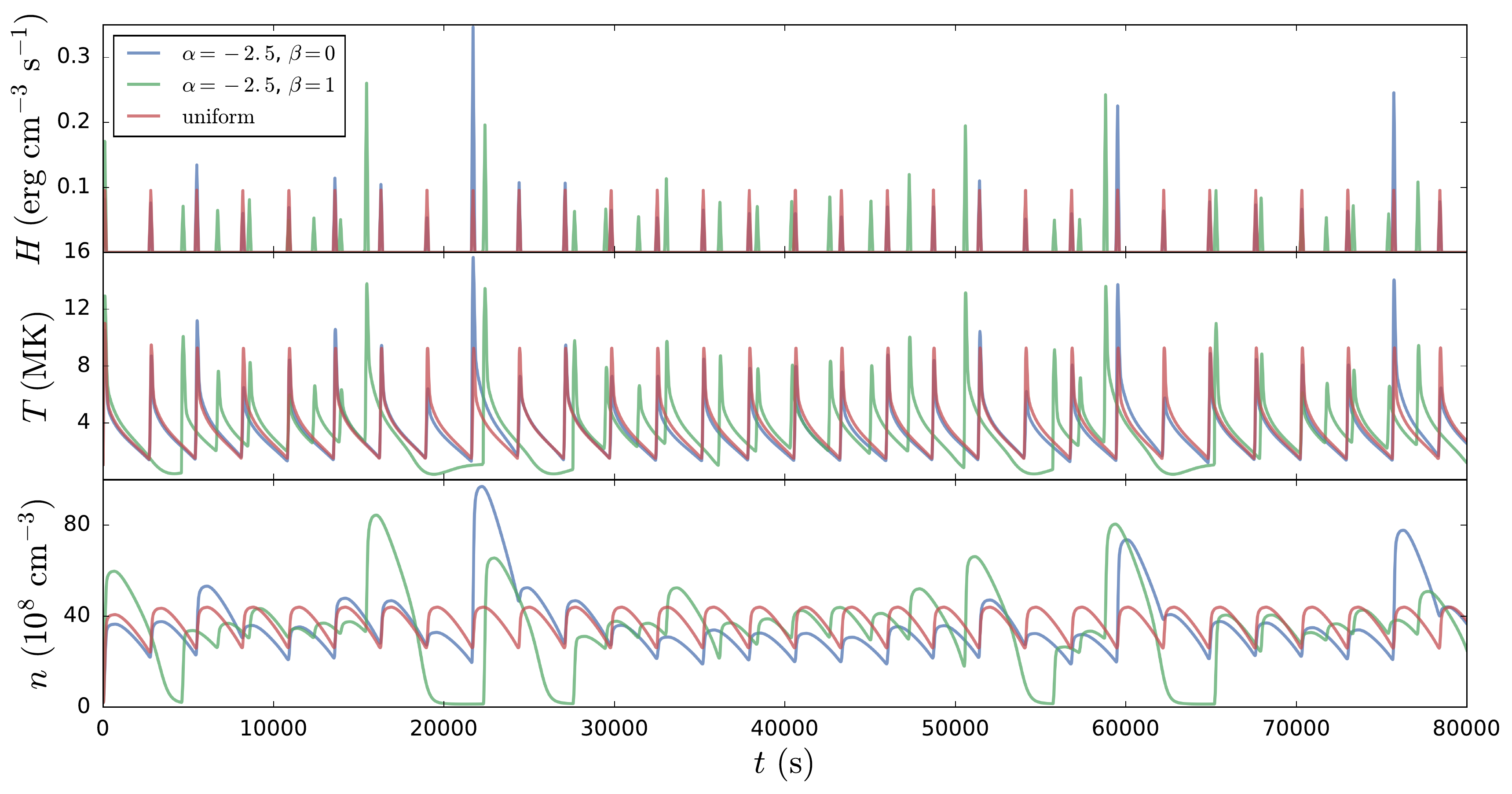}
		\caption{Example heating (top), temperature (middle), and density (bottom) profiles for the case in which only the electrons are heated with a waiting time of $t_N=2500$ s (i.e. an intermediate heating frequency). The three curves shown in each panel correspond to uniform heating rates (red), heating rates chosen from a power-law distribution of $\alpha=-2.5$ (blue), and heating rates chosen from a power-law distribution of $\alpha=-2.5$ where the time between successive events is proportional to the heating rate of the preceding event (green).}
		\label{fig:nT_sample_profiles}
	\end{figure*}
	\par In \citet{cargill_active_2014}, \citetalias{barnes_inference_2016}, and \autoref{subsec:em_dist}, we have suggested that the plasma density prior to the nanoflare occuring is a crucial parameter in determining the emission measure distribution. This arises in two distinct ways. Below $T_m$, the temperature and density at which the nanoflare occurs cuts off the emission at lower temperatures. When combined with an energy-dependent waiting time, this can lead to a range of EM slopes in this region \citep{cargill_active_2014}. Above $T_m$, the initial density determines the temperature increase due to the nanoflare, how quickly the initial hot plasma cools, and whether NEI effects are important. We now examine this further.
	\par In the single-fluid and electron heating cases, while $\mathrm{EM}(T)$ in the uniform and power-law heating cases generally agree for low-frequency heating ($t_N=5000$ s), for intermediate frequencies ($t_N\approx750-2500$ s), the power law cases show an enhanced high-temperature component compared to the uniform case as seen in \autoref{fig:single_em} and \autoref{fig:el_em}. \autoref{fig:nT_sample_profiles} shows sample heating, temperature, and density profiles for an intermediate heating frequency (i.e. a waiting time of $t_N=2500$ s), in the case where only the electrons are heated, for the three different types of heating functions. In the uniform heating rate case (red), each event has a maximum heating rate of $H_0$ such that the loop undergoes $N\approx30$ identical heating and cooling cycles, each time reaching a maximum temperature and density of $T_{max,0}$ and $n_{max,0}$, respectively.
	\par In comparing various heating models, we insist that the total energy injected into the loop is the same for each run (see \autoref{eq:heating_rate_constraint}). When the nanoflare heating rates are distributed according to a power law, there will be many events where $H_i<H_0$ and a few events where $H_i\gg H_0$. These few high energy events lead to $T\gg T_{max,0}$ (blue and green curves) as seen in the middle panel of \autoref{fig:nT_sample_profiles}. Because these events are injected into a plasma that is sufficiently dense due to the draining and cooling times being longer than the time since the previous event, the emission measure is able to ``see'' these hot temperatures, resulting in a $>10$ MK component of $\mathrm{EM}(T)$ (see lower left panel of \autoref{fig:single_em} and \autoref{fig:el_em}). In the uniform case, $T_{max,0}<10$ MK such that $\mathrm{EM}(T)$ has a steep cutoff right at 10 MK.
	\subsection{Hot Plasma Diagnostics}
	\label{subsec:diagnostics}
	\par The relation of these results to potential solar observations is made difficult by incomplete temperature coverage. For example, we noted earlier that the temperature coverage for \textit{Hinode} and SDO is good for $T<T_m$, but patchy for $T>T_m$. On the other hand, the proposed MaGIXS \citep{kobayashi_marshall_2011,winebarger_new_2014} instrument has the opposite performance regime. This differs from EUVE observations of some stellar coronae which have very complete temperature coverage from $10^6$ K to over $10^7$ K \citep[e.g.][]{sanz-forcada_structure_2003}, and have been modeled using low frequency nanoflares \citep{cargill_temperature-emission_2006}. The results in \autoref{subsec:em_dist} suggest that given good spectral resolution, there may be detectable features in $\mathrm{EM}(T)$, although atomic physics and other uncertainties would still be a concern. Instead, in the following sub-sections we propose ways that the paucity of solar temperature coverage can be partially remedied by a consideration of simple metrics.
	\subsubsection{Emission Measure Slope}
	\label{subsubsec:em_slope}
	\begin{figure*}
		\plotone{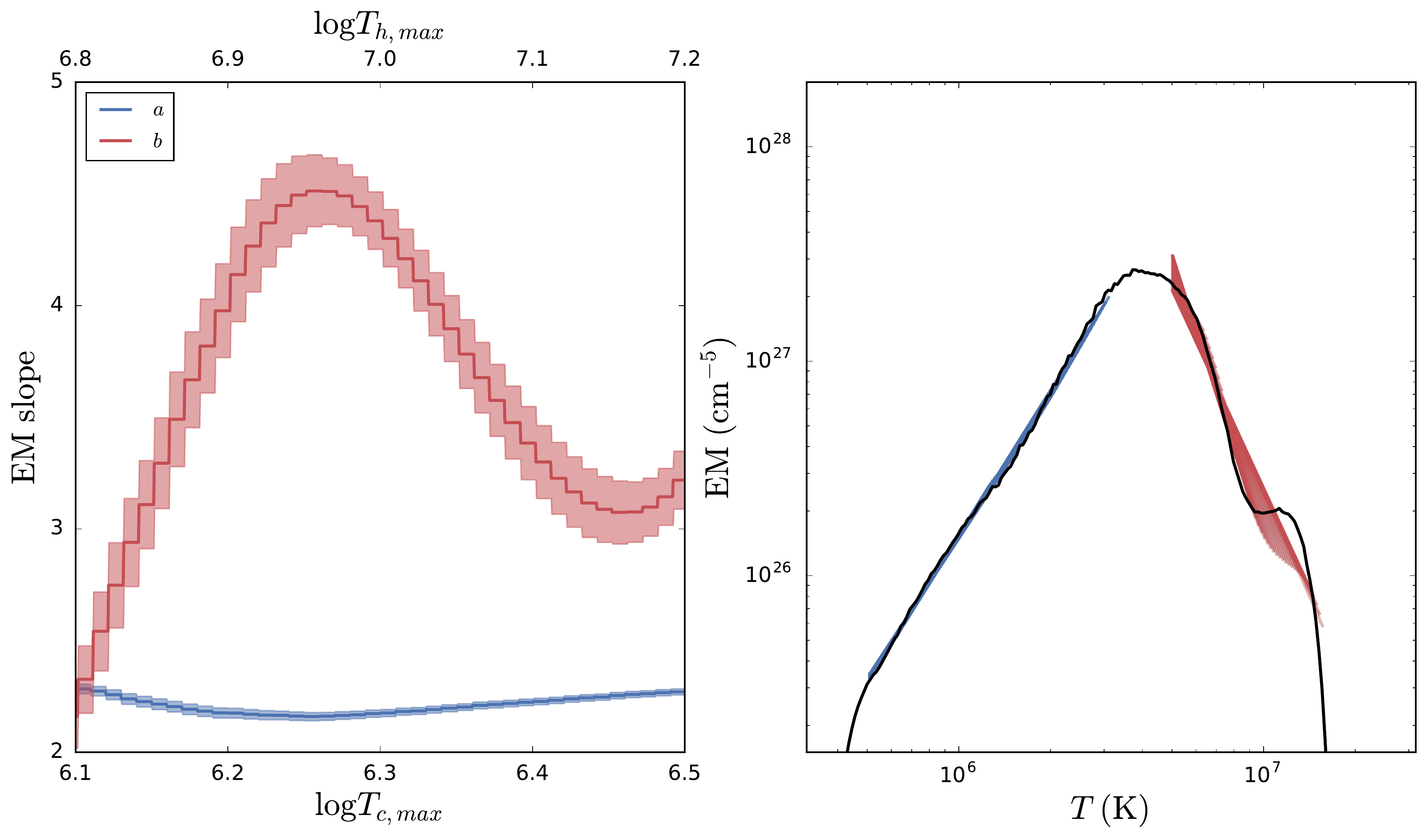}
		\caption{Fits to a sample emission measure distribution constructed from a loop plasma in which only the electrons were heated by events chosen from a power-law distribution with $\alpha=-2.5$ and equally spaced by an interval of $t_N=5000$ s. \textbf{Left:} Emission measure slope as a function of upper bound on the fit interval for both the hot (red) and cool (blue) side of $\mathrm{EM}(T)$. The shading denotes the uncertainty of the fit. The bottom axis corresponds to the varying upper limit on the fit to the cool side while the top axis corresponds to the varying upper limit on the fit to the hot side. \textbf{Right:} $\mathrm{EM}(T)$ with the overlaid hot (red) and cool (blue) fit lines whose slopes correspond to those shown on the left. The cool power-law fits describe $\mathrm{EM}(T)$ for $T<4$ MK quite well while a similar fit on the hot side fails to accurately describe the shape of $\mathrm{EM}(T)$ for $T>4$ MK.}
		\label{fig:em_slope_varying_bounds}
	\end{figure*}
	\par As we discussed in the \autoref{sec:intro}, a commonly-used observable is the emission measure slope $a$ such that $\mathrm{EM}\propto T^a$ for $10^{5.5}\le T\le10^{6.6}$ K. Both observational and modeling studies have found that $2\lesssim a\lesssim5$ \citep[see Table 3 of][]{bradshaw_diagnosing_2012} and in particular, \citet{cargill_active_2014} found that a heating function of the form $t_N\propto\varepsilon$ was needed in order to account for this range of slopes. In this range, the EIS instrument \citep{culhane_euv_2007} on \textit{Hinode} permits good temperature coverage. Additionally, a similar scaling of $\mathrm{EM}\propto T^{-b}$ for $10^{6.6}\le T\le10^{7.0}$ has been claimed though measurements of $b$ have been subject to large uncertainties \citep{warren_systematic_2012} due to intermittent temperature coverage.
	\par\autoref{fig:em_slope_varying_bounds} shows an example of how both $a$ and $b$ can be calculated from the cool and hot sides of $\mathrm{EM}(T)$, respectively. We select a single sample run from our parameter space in which only the electrons are heated by nanoflares from a power-law distribution of $\alpha=-2.5$ and spaced uniformly by an interval of $t_N=5000$ s. We calculate the resulting $\mathrm{EM}(T)$ and fit $\log{\mathrm{EM}}$ to $a\log{T}$ on $\log{T_{c,min}}<\log{T}<\log{T_{c,max}}$ and $-b\log{T}$ on $\log{T_{h,min}}<\log{T}<\log{T_{h,max}}$ using the Levenburg-Marquardt algorithm for least-squares curve fitting. We fix the lower limit on each interval such that $T_{c,min}=10^{5.7}$ K and $T_{h,min} = 10^{6.7}$  K and vary the upper limits over $10^{6.1}<T_{c,max}<10^{6.5}$ K and $10^{6.8}<T_{h,max}<10^{7.2}$ K. The left panel of \autoref{fig:em_slope_varying_bounds} shows $a$ (blue) and $b$ (red) as a function upper limit of the fit interval, $T_{c,max}$ (bottom axis) and $T_{h,max}$ (top axis), respectively. The shading denotes the uncertainty of the fit. The right panel of \autoref{fig:em_slope_varying_bounds} shows the resulting fit lines superimposed on the emission measure distribution.
	\par From the left panel of \autoref{fig:em_slope_varying_bounds}, we see that, while $a$ is relatively insensitive to the fit interval, $b$ varies between approximately 2 and 4.5 depending on the choice of bounds. Furthermore, the uncertainty in the fitting procedure for $b$ is relatively large, with the average uncertainty over the entire range of $T_{h,max}$ being $\bar{\sigma}_b\approx0.17$. Contrastingly, we find that $a\approx2.3$ with little variation over all values of $T_{c,max}$ considered here and that $\bar{\sigma}_a\approx0.018$, nearly an order of magnitude smaller than $\bar{\sigma}_b$. The overlaid fit lines in the right panel of \autoref{fig:em_slope_varying_bounds} similarly show that while $\log{\mathrm{EM}}$ is roughly linear over $5.7<\log{T}<6.5$, this is not the case for the interval $6.7<\log{T}<7.2$. In particular, a function of the form $T^{-b}$ cannot describe the hot shoulder in the emission measure distribution near $10^{7.1}$ K.
	\par Our results here suggest that while $a$ is an adequate parameter for describing the cool side of $\mathrm{EM}(T)$, the functional form $\mathrm{EM}\sim T^{-b}$ does not adequately capture the character of the hot part of $\mathrm{EM}(T)$ over any reasonable temperature range. \citet{antiochos_evaporative_1978} showed analytically that accounting for evaporative cooling and assuming constant pressure gives $b=11/2$, though this value can be as low as $b=5/2$ if a flux limiter is included. However, in \citetalias{barnes_inference_2016}, we showed that in the case of a single 200 s nanoflare that heats only the electrons, the assumption of constant electron pressure during the heating and early conductive cooling phases does not hold. Our results here are consistent with our findings in \citetalias{barnes_inference_2016} in that the parameter $b$ does not provide any valuable information about $\mathrm{EM}(T)$ for $T>T_m$ when two-fluid effects are considered. Clearly, an alternative metric for measuring the amount of hot plasma in the emission measure distribution is needed.
	\subsubsection{Emission Measure Ratio}
	\label{subsubsec:em_ratio}
	\begin{figure*}
		\plottwo{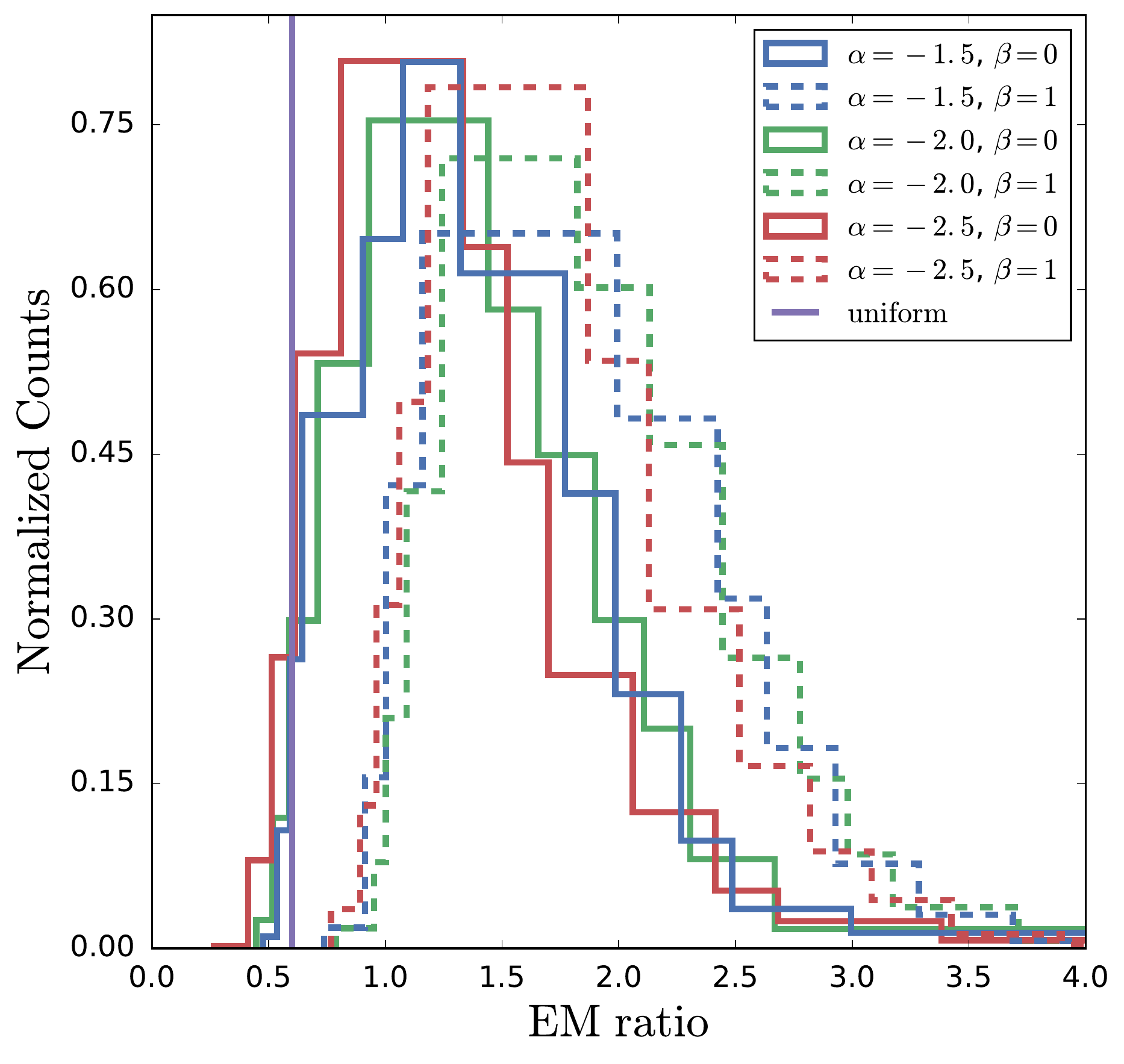}{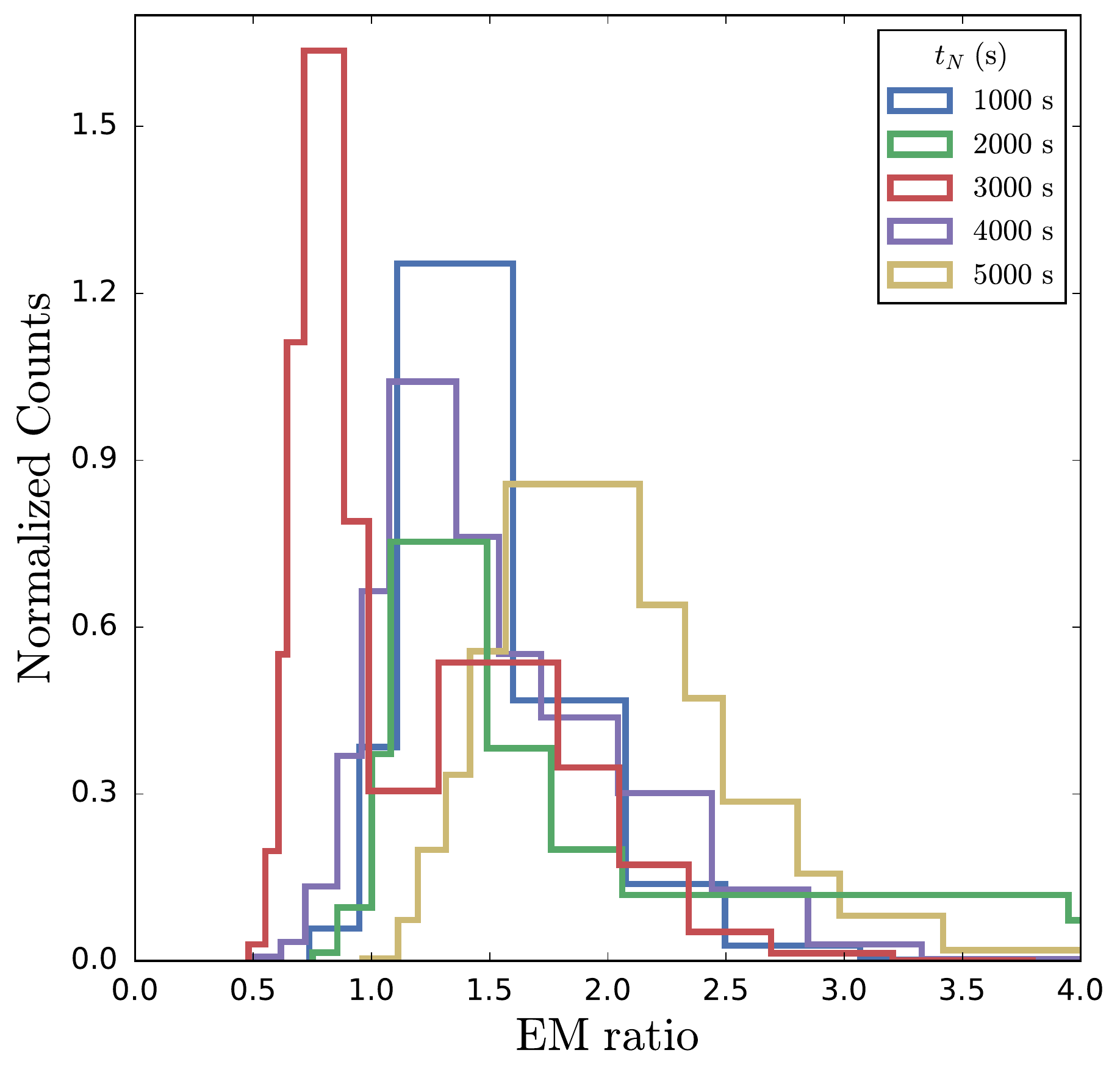}
		\caption{Histograms of emission measure ratios $\mathrm{EM}(T_{hot})/\mathrm{EM}(T_{cool})$, where $T_{hot}=10^{6.942}$ K and $T_{cool}=10^{6.187}$ K, for all heating function types and heating frequencies for the single fluid case. In both panels each histogram is normalized such that for each distribution $P(x)$, $\int_{-\infty}^{\infty}\mathrm{d}x~P(x)=1$ and the bin widths are calculated using the Bayesian blocks method of \citet{scargle_studies_2013}. \textbf{Left:} Emission measure ratios separated by heating function type for all heating frequencies, $250\le t_N\le5000$ s. Because there are too few ($<20$) $\mathrm{EM}$ ratio measurements for the uniform case to construct a meaningful histogram, we denote the median of the uniform results with a vertical line, shown here in purple. \textbf{Right:} Emission measure ratios separated by waiting time, $t_N$. For aesthetic purposes, only five values of $t_N$ are shown, $t_N=1000,2000,3000,4000,5000$ s.}
		\label{fig:em_ratio_hist_single}
	\end{figure*}
	\begin{figure*}
		\plottwo{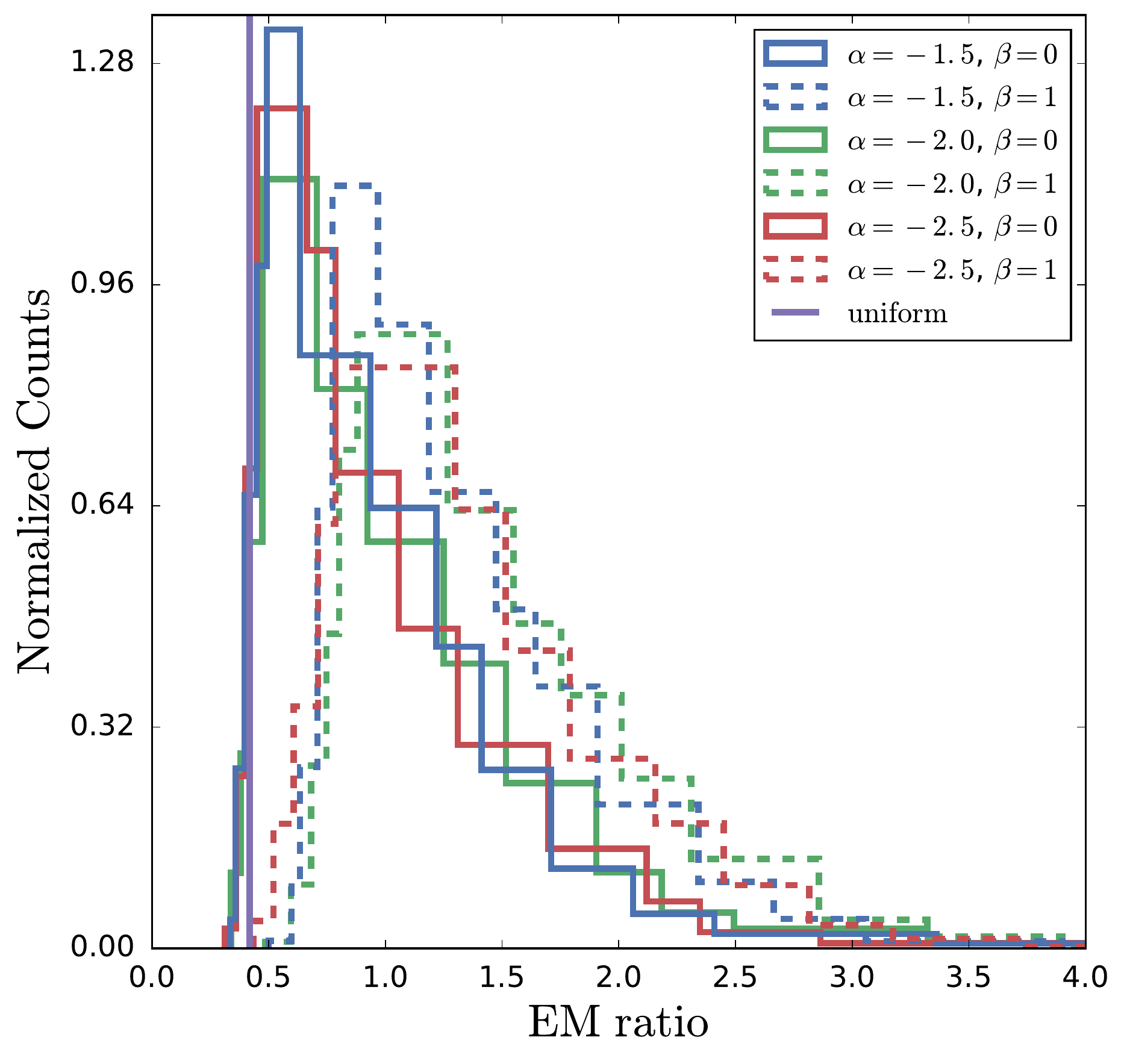}{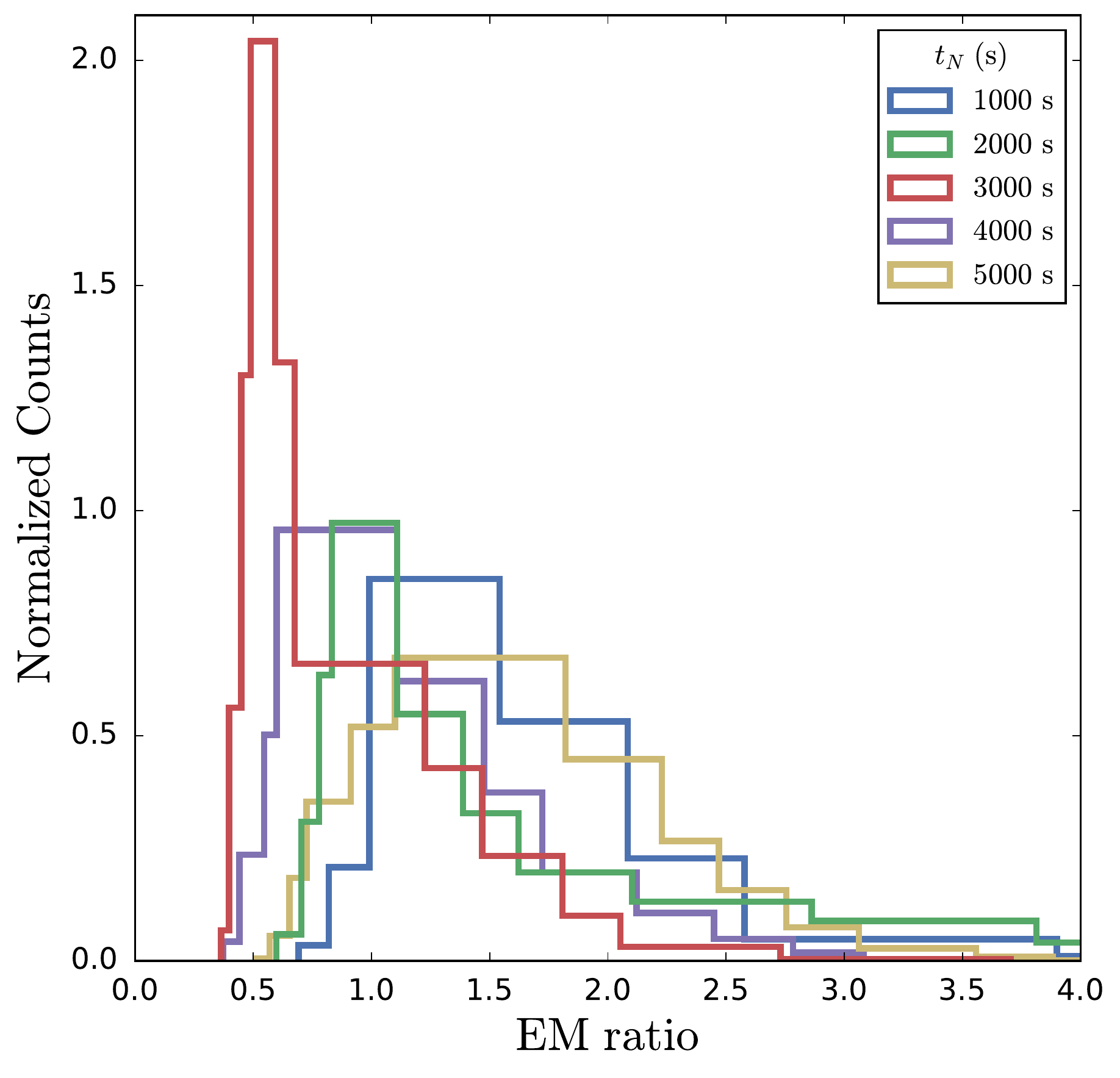}
		\caption{Same as \autoref{fig:em_ratio_hist_single}, but for the case where only the electrons are heated.}
		\label{fig:em_ratio_hist_el}
	\end{figure*}
	\begin{figure*}
		\plottwo{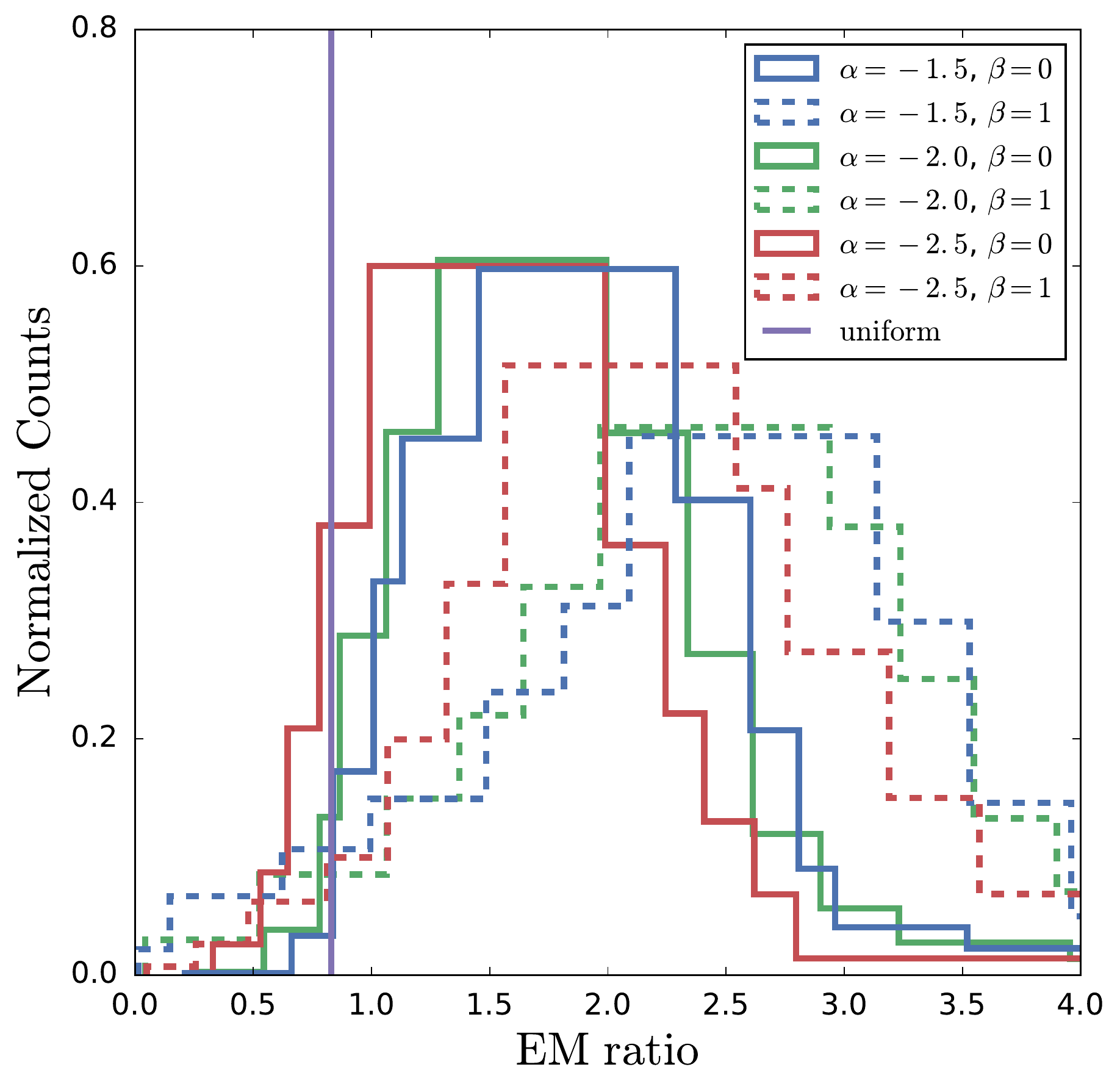}{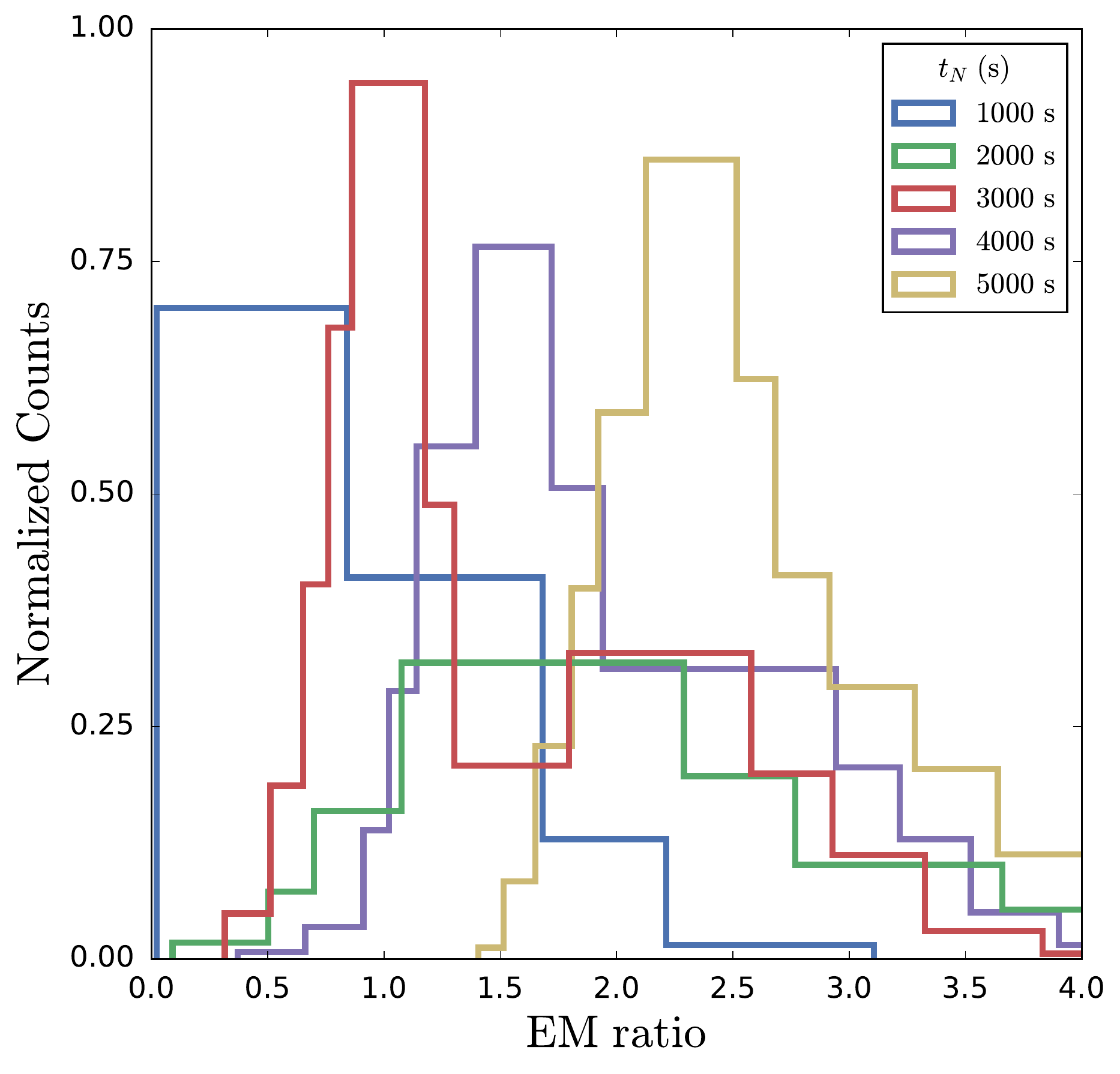}
		\caption{Same as \autoref{fig:em_ratio_hist_single}, but for the case where only the ions are heated.}
		\label{fig:em_ratio_hist_ion}
	\end{figure*}
		\par\citet{brosius_pervasive_2014} proposed another possible diagnostic for the hot non-flaring corona. Using observations of an active region from the \textit{EUNIS-13} sounding rocket, they found that the intensity ratio of Fe XIX (formed at $T\approx8.9$ MK) to Fe XII (formed at $T\approx1.6$ MK) is $\sim0.59$ inside the AR core as compared to $\sim0.076$ outside. They argue that this provides possible evidence for impulsive heating. This arises from the fact that the $\mathrm{EM}(T)$ distribution resulting from low- and intermediate-frequency nanoflares is dual-valued as a function of temperature \citep[see \autoref{fig:single_em}-\autoref{fig:ion_em} and][]{cargill_implications_1994}. For example, in the lower right panel of \autoref{fig:single_em}, $\mathrm{EM}\approx 10^{26.5}$ cm$^{-5}$ at $T = 1.6$ MK and $8.9$ MK. This suggests that the approach of \citet{brosius_pervasive_2014} could be used for any pair of appropriate emission lines and is likely to be of particular use where very limited spectral coverage is available either above or below $T_m$. Indeed for impulsive heating of any kind, the generic $\mathrm{EM}(T)$ curve remains similar (though not identical) as the magnitude of the heating changes. For flare-like energy release one would expect the hot emission to be much greater than the cold emission, with the opposite in the quiet Sun. It should be noted that there can be considerable uncertainties in deriving a reliable $\mathrm{EM}(T)$ distribution from observations due to uncertainties from the atomic physics, line-of sight, loop geometry, etc., which will hopefully be mitigated by improved knowledge of such quantities in the future. \citet{guennou_can_2013} provide an extensive overview of this topic.
	\par We define a general emission measure ratio, $\mathrm{EM}(T_{hot})/\mathrm{EM}(T_{cool})$ and for this paper consider $T_{hot}=10^{6.942}$ K and $T_{cool}=10^{6.187}$ K, the formation temperatures of Fe XIX and XII, respectively \citep[i.e. the temperature $T$ which maximizes the contribution function $G_{\lambda}(T)$ calculated using CHIANTI v8,][]{dere_chianti_1997,del_zanna_chianti_2015}. This also provides a way to compare in a concise way every point in our multidimensional parameter space though we acknowledge that we are reducing $8\times10^4$ s of loop evolution to a single value. This ratio is shown in \autoref{fig:em_ratio_hist_single}, \autoref{fig:em_ratio_hist_el}, and \autoref{fig:em_ratio_hist_ion} for the single-fluid model, electron and ion heating, respectively. In the left-hand panels each individual histogram (denoted by linestyle and color) corresponds to a different type of heating function. This means, for example, that the solid blue histogram includes emission measure ratios for all values of $t_N$, but for only those cases where heating events are evenly spaced (i.e. $\beta=0$) and chosen from a power-law distribution of $\alpha=-1.5$. The right panels show these same emission measure ratios, but now categorized by $t_N$. For example, the solid red histogram includes emission measure ratios for every type of heating function (i.e. uniform, all $\alpha$ and all $\beta$), but for only those runs where $t_N=3000$ s. Note that we choose to only show results for five values of $t_N$ for aesthetic purposes.
	\par Considering first the left hand panels, in the single fluid model (\autoref{fig:em_ratio_hist_single}) the ratio is largely insensitive to $\alpha$ and is peaked sharply at $\sim1-1.25$ although the distribution peaks at slightly higher values for the $\beta=1$ case ($\sim1.5$). Note that the uniform heating results (whose median is denoted by the vertical purple line near $\sim0.6$) show emission measure ratios significantly less than those in the power law cases, consistent with the reduced hot emission shown in \autoref{fig:single_em}. For electron heating (\autoref{fig:em_ratio_hist_el}) the ratio is again insensitive to $\alpha$, but is narrower and falls off more quickly toward higher emission measure ratios. The $\beta=0$ cases (for all $\alpha$) peak just above $0.5$, with the $\beta=1$ cases all peaking at slightly higher values just at or below 1. For ion heating (\autoref{fig:em_ratio_hist_ion}) the $\beta=0$ distributions peak at lower values compared to those with $\beta=1$. Again the results are relatively insensitive to $\alpha$. However, compared to the electron heating case, all of the distributions are much wider and peak at higher values, $\sim1.75$ for $\beta=0$ and $\sim2-2.5$ for $\beta=1$. Furthermore, in the $\beta=1$ case, the $\alpha=-2.5$ distribution peaks at lower values compared $\alpha=-1.5,-2.0$.
	\par Turning to the right hand panels, in the single-fluid model for $t_N\le4000$ s, the results cluster just near 1.5, though the $t_N=3000$ s distribution is slightly bimodal, peaking strongly at $\sim0.75$ and more weakly at $\sim1.5$. Additionally, the $t_N=5000$ s case peaks slightly higher at $\sim2$ and does not ``pile up'' near $1.5$ as the other values of $t_N$ do. For electron heating, the $t_N=3000$ s distribution has a very strong and narrow peak at $\sim0.5$ while all of the other distributions peak at $\sim1-1.5$. For $t_N\ge3000$ s, the location of the peak increases weakly with increasing $t_N$ though this is not the case for $t_N=1000,2000$ s. Finally, for ion heating the distributions for each value of $t_N$ are much wider than for the single fluid and electron heating cases and the peak values show a stronger dependence on $t_N$. The range of peak values is also much larger, with the $t_N=1000$ s case peaking near 0 and the $t_N=5000$ s case peaking just at $\sim2.5$. As in \autoref{fig:em_ratio_hist_single}, we see that the $t_N=3000$ s distribution is bimodal. In general, the distributions grouped by $t_N$ for all three heating types have a positive skew and are peaked in the range $\sim0.5-2$ except for some extreme cases in the ion heating scenario.
	\par These results suggest that the emission measure ratio is generally in the range $\sim0.5-2$, with some higher values. Given the uncertainties in the atomic physics, this seems to support the conclusion of the presence of nanoflare heating by \citet{brosius_pervasive_2014}. Thus the calculation of such ratios from limited data has the potential to be a powerful diagnostic of the existence of nanoflare heating. Whether one can say more about the precise form of nanoflare heating is less clear. In these results the ratios are largely independent of the power-law index $\alpha$ and only weakly-dependent on the waiting time, $t_N$. Furthermore, the distributions are also weakly dependent on the relationship between the waiting time and heating rate, $\beta$. The exception is the ion heating case where the distributions are much wider, peak at higher values, and show a stronger dependence on $\alpha$, $\beta$, and $t_N$. The problem with drawing conclusions from the details of these results lies in the numerous uncertainties in any data, especially concerning atomic physics.
	\subsubsection{Additional Line Pairs}
	\label{subsubsec:add_line_pairs}
	\begin{figure*}
		\plotone{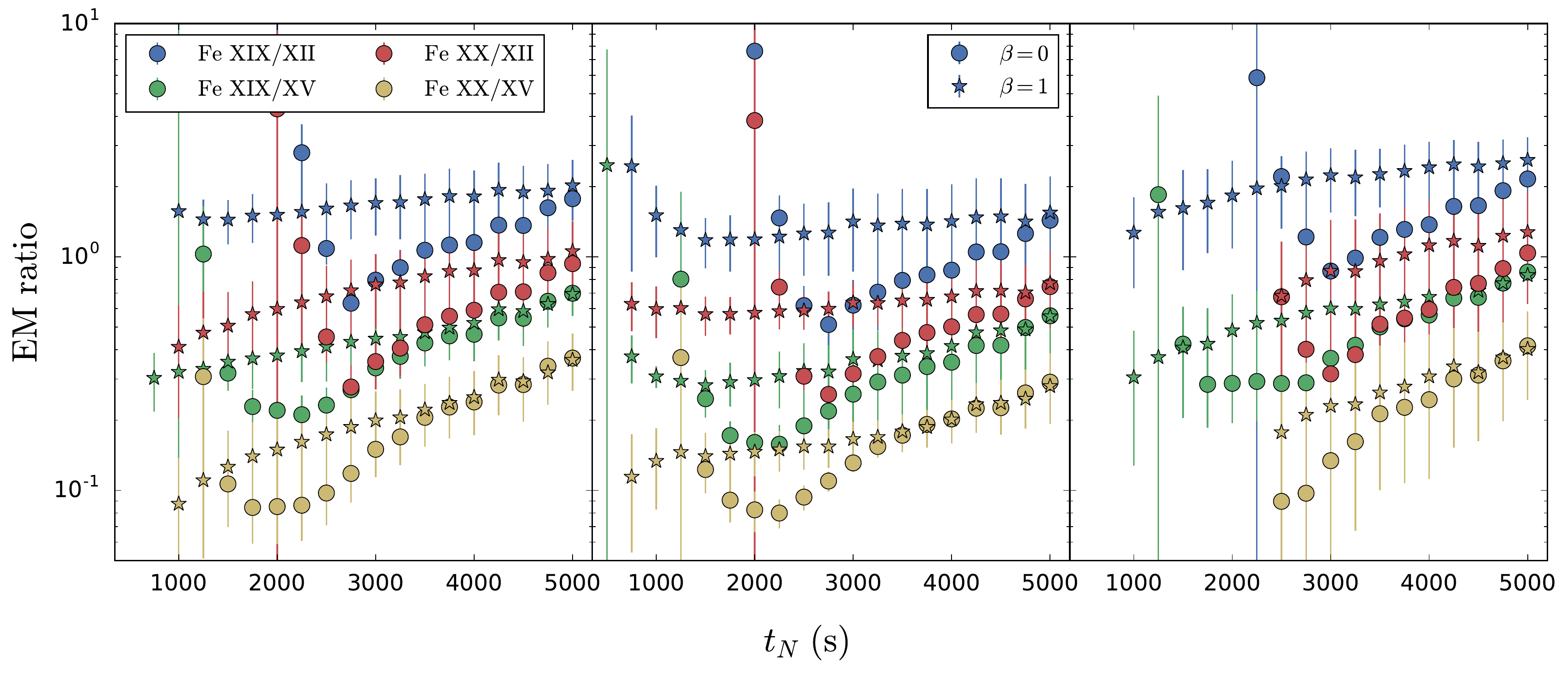}
		\caption{Emission measure ratio as a function of $t_N$ for the single-fluid (left), electron heating (center), and ion heating (right) cases for four different line pairs: Fe XIX/Fe XII (blue), Fe XIX/Fe XV (green), Fe XX/Fe XII (red), and Fe XX/Fe XV (yellow). The circles ($\beta=0$) and stars ($\beta=1$) indicate the mean EM ratio calculated over all runs for a given $t_N$ (see \autoref{subsec:heating_stats}) and the error bars indicate one standard deviation from the mean. Here we only show the results for $\alpha=-2.5$.}
		\label{fig:em_ratio_vs_tn}
	\end{figure*}
	\par Because our proposed EM ratio diagnostic is based on only two points from each $\mathrm{EM}(T)$, it is reasonable to ask how this measurement might be affected by a different choice of hot/cool temperatures. \autoref{fig:em_ratio_vs_tn} shows the mean EM ratio (with the error bars corresponding to one standard deviation) as a function of $t_N$ for three additional line pairs: Fe XIX/XV, Fe XX/XII, and Fe XX/XV, as denoted by the color coding. The formation temperatures of Fe XV and Fe XX, calculated using CHIANTI v8, are $10^{6.342}$ K and $10^{7.027}$ K, respectively. The circles (stars) correspond to $\beta=0(1)$ and we consider only $\alpha=-2.5$ here. The values of the line ratios can be organized from high to low as Fe XIX/Fe XII, Fe XX/Fe XII, Fe XIX/Fe XV and Fe XX/Fe XV. The second and third pairs have similar ratio values. This ordering is to be expected from the general shape of the $\mathrm{EM}(T)$ curves, although the exact ratios cannot be predicted in any simple way.
	\par For $\beta=1$, the ratio decreases monotonically as $t_N$ decreases, however the values of the ratio lie within a narrow band. Given the many uncertainties in the measurements, it is unlikely that any of these line pairs can provide an unequivocal constraint on $t_N$. For $\beta=0$, there is an upturn in the ratio between $t_n = 2000$ s and $3000$ s, and below this value, the ratios are ill-defined due to lack of plasma at one or both of the measured temperatures. At lower temperatures, this arises because, for an evenly spaced nanoflare train, there is a sharp cut-off in $\mathrm{EM}(T)$ at some temperature $T< T_m$ which does not occur when a waiting time is included \citep{cargill_active_2014}.
	\par On the other hand, the relative ratios of line pairs is a quantitative prediction that can be examined were such data to be available, and can be easily extended to other line pairs not considered here. In addition, for all line pairs, the detection of an emission measure ratio lying between the upper and lower bounds shown can be regarded as a very strong indicator of the presence of ``hot'' nanoflare-produced plasma. Further deductions require more detailed modeling of in particular the atomic physics which is beyond the scope of this paper.
	\section{Conclusions}
	\label{sec:conclusions}
	\par In this paper we have carried out two-fluid modeling of nanoflare trains in AR cores and considered a range of models for the nanoflare energy distribution and timing as well as preferential heating of different species. For each set of parameters we have generated the emission measure distribution as a function of temperature. If the peak of the emission measure occurs at $T = T_m$, then we found that $T_m$ and $\mathrm{EM}(T_m)$ were independent of the properties of the nanoflare train and which species was heated. As a consequence, we demonstrated that the form of $\mathrm{EM}(T)$ on either side of $T_m$ reflected different aspects of the heating process. Below $T_m$ the principle factor in determining $\mathrm{EM}(T)$ is the presence of a waiting time between nanoflares: the high densities below $T_m$ mean that species temperature equilibration has occurred so that no information about which species was heated remains.
	\par Above $T_m$, higher temperatures arise when a power-law distribution of energy is assumed than for a nanoflare train with uniform heating rates. However, no information about the presence of a waiting time survives. Higher observed temperatures also arise for the single-fluid and electron heating cases where a ``hot shoulder'' can occur in $\mathrm{EM}(T)$ at $T\sim10$ MK. This is compared to the ion heating case where a finite equilibration time means that the electrons are heated more slowly. We find that, unlike some of the examples in \citetalias{barnes_inference_2016}, NEI is not a major consideration, a constant nanoflare train being the exception.
	\par Two possible ways to relate these results to present and future observations were discussed. First we showed that while below $T_m$ the well-known relation $\mathrm{EM}\sim T^a$ was quite robust, that was not the case above $T_m$. On the other hand the calculation of a ratio between the emission measure at a pair of temperatures above and below $T_m$ held more promise. For a wide range of parameters the ratio of the emission measure at two temperatures, $T_{hot}=10^{6.942}>T_m$ and $T_{cool}=10^{6.187}<T_m$, was of order unity, consistent with \citet{brosius_pervasive_2014} within various errors. Further observational evidence of such ratios would, in our view, provide a very strong case for the presence of nanoflare heating. On the other hand, given the uncertainties in atomic physics and emission emasure analysis, such a comparison seems unlikely to be able to shed more detailed light on the details (e.g. $t_N$, $\alpha$, $\beta$) of the actual heating process.
	\par It is clear from this and the preceeding paper that characterization of this ``hot'' component is extremely challenging for a multiplicity of reasons. Progress is likely to come from the good spectral coverage of the MaGIXS instrument, and in particular from a space-based successor which could confirm beyond doubt the presence of the hot component and perhaps measure the predicted features of $\mathrm{EM}(T)$. In the absence of complete spectral coverage, we propose a pair of metrics for the ``hot'' coronal component. Of particular interest is the ratio of pairs of emission lines characteristic of cool and hot plasma, as was recently discussed by \citet{brosius_pervasive_2014}. When high temperature spectral coverage is limited, information from high-energy instruments \citep{ishikawa_constraining_2014,hannah_first_2016,grefenstette_first_2016} would be then desirable, but the energies of interest ($\approx1$ keV) are highly challenging. In any event, complete wavelength coverage seems essential, something that we noted stellar astronomers have had for decades.
	\acknowledgments
	WTB was provided travel support to the Coronal Loops Workshop VII held in Cambridge, UK, at which a preliminary version of this work was presented, by NSF award number 1536094. This work was supported in part by the Big-Data Private-Cloud Research Cyberinfrastructure MRI-award funded by NSF under grant CNS-1338099 and by Rice University.
	\software{astroML \citep{vanderplas_introduction_2012}, IPython/Jupyter \citep{perez_ipython:_2007}, matplotlib \citep{hunter_matplotlib:_2007}, NumPy/scipy \citep{van_der_walt_numpy_2011}, seaborn \citep{waskom_seaborn:_2016}}
	%
	%
	\bibliography{references.bib}
	\bibliographystyle{aasjournal}
\end{document}

%% file: parameter_space_graphic.tex
	\begin{tikzpicture}[node distance=2cm]
		\node (species) [ghost] {Heated Species $=\left\{
		\begin{array}{l}
			\mathrm{electron} \\
			\mathrm{ion} \\
			\mathrm{single}
		\end{array}
		\right.$};
		\node (ghost_ph) [ghost, below of=species] {};
		\node (alpha_pl) [ghost, left of=ghost_ph] {$\alpha=\left\{
		\begin{array}{l}
			-1.5 \\
			-2.0 \\
			-2.5
		\end{array}
		\right.$};
		\node (alpha_uni) [ghost, right of=ghost_ph] {uniform};
		\node (ghost_beta) [ghost, below of=alpha_pl]{};
		\node (beta_1) [ghost, right of=ghost_beta] {$\beta=1$};
		\node (beta_0) [ghost, left of=ghost_beta] {$\beta=0$};
		\draw [arrow] (species) -- (alpha_pl);
		\draw [arrow] (species) -- (alpha_uni);
		\draw [arrow] (alpha_pl) -- (beta_0);
		\draw [arrow] (alpha_pl) -- (beta_1);
	\end{tikzpicture}